\documentclass[aps,twocolumn,showpacs]{revtex4}
\usepackage{epsfig}
\usepackage{amsbsy}
\usepackage{amsmath}
\usepackage{amsfonts}
\usepackage{graphicx}
\usepackage{latexsym}

\begin{document}

\title{Entanglement and spin squeezing in the three-qubit transverse Ising model}
\author{XiaoGuang Wang}
\affiliation{Zhejiang Institute of Modern Physics, Department of
Physics, Zhejiang University, HangZhou 310027, China}

\begin{abstract}
We study entanglement and spin squeezing in the ground state of
three qubits interacting via the transverse Ising model. We give
analytical results for the entanglement and spin squeezing, and a
quantitative relation between the concurrence, quantifying the
entanglement of two spins, and the spin squeezing parameter,
measuring the degree of squeezing. Finally, by appropriately
choosing the exchange interaction and strengths of the transverse
field, we propose a scheme for generating entangled W state from
an unentangled initial state with all spins down.
\end{abstract}
\maketitle

Entanglement in many-body systems have attracted much
attention~\cite{Nielsen}-\cite{GVidal}. Spin squeezing, resulting
from quantum correlations between spins, has been a subject of
much studies in recent years~\cite{Wodkiewicz}-\cite{Rojo}.
Interestingly, there exist close relations between the
entanglement and spin squeezing~\cite{Sorensen1, WangBarry}, which
enhances the importance of spin squeezing.

In this paper, we study ground-state entanglement and spin
squeezing in the three-qubit transverse Ising model (TIM), which
can be exactly solved. For the two-qubit case, analytical results
of the entanglement in the TIM have been given by Gunlycke et
al.~\cite{Gunlycke}. Here, we give analytical results of pairwise
entanglement and spin squeezing for the three-qubit case. It was
found that the spin squeezing is equivalent to bipartite
entanglement for an arbitrary two-qubit pure state~\cite{Ulam}.
However, for the case of three qubits or more, in general the
entanglement is not equivalent to spin squeezing, and there could
exist some relations between them~\cite{WangBarry}. We study
relations between entanglement and spin squeezing for three-qubit
states. We also propose a scheme for dynamically producing an
entangled W state~\cite{Dur,Eibl,Marr,Guo1,Guo2,Wang01} in the
TIM.

{\em Transverse Ising model.} We first consider the $N$-qubit TIM,
and the corresponding Hamiltonian is given by
\begin{equation}
H=J\sum_{i=1}^N \sigma _{ix}\otimes\sigma _{i+1x}+
B\sum_{i=1}^N \sigma _{iz},
\end{equation}
where $\vec{\sigma}_i=(\sigma _{ix},\sigma _{iy},\sigma _{iz})$ is
the vector of Pauli matrices, $J$ is the exchange constant, and
$B$ is the magnetic field. The positive and negative $J$
correspond to the antiferromagnetic (AFM) and ferromagnetic (FM)
case, respectively. We assume periodic boundary conditions, i.e.,
$N+1\equiv 1$, and $J>0$, $B>0$.

There are several symmetries in the TIM which are useful for our
analysis. The first obvious symmetry of $H$ are is the translation
invariance, which implies that the entanglement between any
nearest qubits are equal. The second one is the $Z_2$ symmetry,
i.e., $H$ commutes with $\Sigma _z,$ where
\begin{equation}
\Sigma _\alpha =\sigma _{1\alpha}\otimes \sigma _{2\alpha }\otimes
\cdot \cdot \cdot \otimes \sigma_{N\alpha }(\alpha =x,y,z).
\end{equation}
This symmetry guarantees that the reduced density matrix of two
nearest qubits , say qubit 1 and 2, for the ground state
$\rho=|\Psi_0\rangle\langle\Psi_0|$ has the following form
\begin{equation}
\rho_{12}=\left(
\begin{array}{llll}
u & 0 & 0 & y \\
0 & w & z & 0 \\
0 & z & w & 0 \\
y & 0 & 0 & v
\end{array}
\right)  \label{rho12}
\end{equation}
in the standard basis $\{|00\rangle ,|01\rangle ,|10\rangle
,|11\rangle \}.$ The reality of $y$ and $z$ results from the third
symmetry, the reflection symmetry, i.e., the Hamiltonian is
invariant under the transformation $\prod_{n=1}^{N/2}S_{n,N-n+1},$
where $S_{ij}$ is the swap operator for qubit $i$ and $j.$ Then,
by the standard procedure for calculating the concurrence
$C$~\cite{Conc}, the entanglement measure of two qubits, one finds
\begin{equation}
C=2\max(0,|z|-\sqrt{uv},|y|-w). \label{conc}
\end{equation}
The fourth symmetry is that the Hamiltonian is invariant under the
joint transformation $B\rightarrow -B$ and $\Sigma _xH\Sigma _x.$

{\em Eigenvalue problem.} Now we restrict ourselves to the
three-qubit case, and solve the eigenvalue problem of the TIM.
Once the ground state is obtained, we can analyse their
entanglement and squeezing properties. Since we impose the
periodic boundary condition, the translational invariance ensures
that the Hamiltonian matrix reduces to submatrices by a factor of
the number of qubits~\cite{Linhq}.  We choose the following basis,
\begin{align}
|\psi_0\rangle&=|000\rangle,\nonumber\\
|\psi_1\rangle&=\frac{1}{\sqrt{3}}(|100\rangle+\omega^2|010\rangle+\omega|001\rangle),
\nonumber\\
|\psi_2\rangle&=\frac{1}{\sqrt{3}}(|100\rangle+\omega|010\rangle+\omega^2|001\rangle),\nonumber\\
|\psi_3\rangle&=\frac{1}{\sqrt{3}}(|100\rangle+|010\rangle+|001\rangle),\nonumber\\
|\psi_4\rangle&=\frac{1}{\sqrt{3}}(|011\rangle+\omega^2|101\rangle+\omega|110\rangle),\nonumber\\
|\psi_5\rangle&=\frac{1}{\sqrt{3}}(|011\rangle+\omega|101\rangle+\omega^2|110\rangle),\nonumber\\
|\psi_6\rangle&=\frac{1}{\sqrt{3}}(|011\rangle+|101\rangle+|110\rangle),\nonumber\\
|\psi_7\rangle&=|111\rangle,
\end{align}
where $\omega=\exp(i2\pi/3)$.
These are all eigenstates of the cyclic right shift operator $T$, which is defined by its action on the product basis
\begin{equation}
T|m_1,m_2,m_3\rangle=|m_3,m_1,m_2\rangle.
\end{equation}
For instance,
\begin{equation}
T|\psi_1\rangle=\omega|\psi_1\rangle.
\end{equation}

By using the above basis, and considering both the translational
invariance and the $Z_2$ symmetry, we see that the $8\times 8$
Hamiltonian matrix of the TIM can be reduced to submatrices that
are no more than $2\times 2$. Thus, all eigenvalues and
eigenstates can be obtained analytically. The eigenvalues are
given by
\begin{align}
E_0=&J-B-\sqrt{3J^2+(J+2B)^2}, \nonumber\\
E_1=&J-B+\sqrt{3J^2+(J+2B)^2}, \nonumber\\
E_2=&J+B-\sqrt{3J^2+(J-2B)^2}, \nonumber\\
E_3=&J+B+\sqrt{3J^2+(J-2B)^2}, \nonumber\\
E_4=&E_5=B-J, \nonumber\\
E_6=&E_7=-B-J.
\end{align}
As we are interested in the ground-state properties, only
ground-state vector is given as follows
\begin{align}
|\Psi_0\rangle=&a_1|111\rangle+a_2|W\rangle \nonumber\\
=&a_1|111\rangle+\frac{a_2}{\sqrt{3}}(|100\rangle+|010\rangle+|001\rangle),\nonumber\\
a_1=&\frac{\sqrt{3}J }{\sqrt{3J^2+(E_0+3B)^2}},\nonumber\\
a_2=&\frac{E_0+3B}{\sqrt{3J^2+(E_0+3B)^2}},\label{gs}
\end{align}
where
\begin{equation}
|W\rangle \equiv |\psi_3\rangle=\frac 1{\sqrt{3}}\left(
|100\rangle +|010\rangle +|001\rangle\right)
\end{equation}
is the W state~\cite{Dur}. Then, the eigenvalue problem is
completely solved, and specifically the ground state is explicitly
obtained. We next study the entanglement and squeezing properties
of the ground state.

{\em Entanglement and spin squeezing.} To study entanglement and
squeezing, we consider a more general state given by
Eq.~(\ref{gs}) with the coefficients $a_1$ and $a_2$ being
arbitrary. Without loss of generality, these coefficients can be
chose as
\begin{equation}
a_1=\cos\theta, a_2=\sin\theta\exp(i\phi),
\end{equation}
where $\phi$ is the relative phase between state $|111\rangle$ and
the W state.

Tracing out the third qubit, we obtain the two-qubit reduced density matrix of
state for qubits 1 and 2, which is given by Eq.~(\ref{rho12}) with the matrix elements
\begin{equation}
y=\frac{a_2a_1^*}{\sqrt{3}},\quad z=w=u=\frac{|a_2|^2}{3}, \quad v=|a_1|^2.
\end{equation}
Thus, from Eq.~(\ref{conc}), the concurrence for qubits 1 and 2 is
obtained as
\begin{equation}
{C}=\frac{2|a_2|}{3}||a_2|-\sqrt{3}|a_1||. \label{ccc}
\end{equation}
We see that the entanglement between qubits 1 and 2 is independent
on the relative phase $\phi$, and only determined by one single
parameter $\theta$.

\begin{figure}
\includegraphics[width=0.45\textwidth]{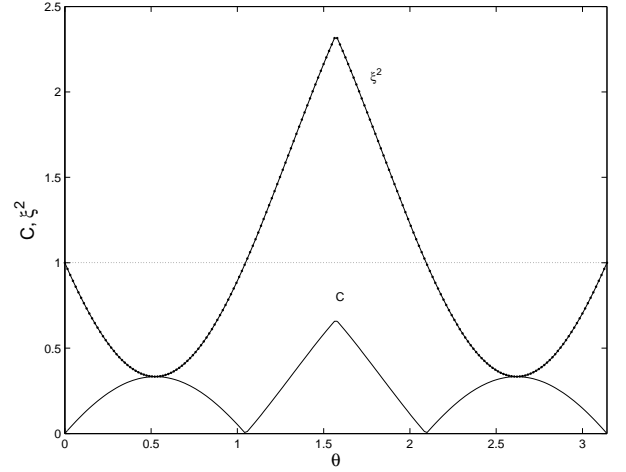}
\caption{The concurrence and spin squeezing parameter versus $\theta$.}
\end{figure}

Now, we consider spin squeezing of the state, which is obviously
of even parity. For the three-qubit state with parity, the spin
squeezing parameter is given by \cite{WangBarry}
\begin{equation}
\xi^2=\frac{5}{2}-\frac{2}{3}[\langle S_z^2\rangle+|\langle S_-^2\rangle|],
\end{equation}
where
\begin{equation}
S_\alpha=(\sigma_{1\alpha}+\sigma_{2\alpha}+\sigma_{3\alpha})/2,
\alpha\in\{x,y,z\},
\end{equation}
and $S_-=S_x-iS_y$. Applying the above equation to the general
state, we obtain
\begin{equation}
\xi^2=1+\frac{4}{3}|a_2|(|a_2|-\sqrt{3}|a_1|). \label{xixixi}
\end{equation}
The squeezing parameter is also independent of the relative phase $\phi$.
We also note that the concurrence and the squeezing parameter is a periodic function of $\theta$ with period $\pi$.

In Fig.~1, we plot the concurrence and the squeezing parameter
versus $\theta$ within one period. There exist two special values
of $\theta$, $\theta_1=\pi/3$ and $\theta_2=2\pi/3$, at which the
concurrence is zero and the squeezing parameter is one. We observe
that for $\theta\in (0,\theta_1)\cup(\theta_2,\pi)$, the
entanglement and spin squeezing are equivalent, i.e., the
entanglement implies spin squeezing and vice versa. For
$\theta\in(\theta_1,\theta_2)$, the state is pairwise entangled,
but not spin squeezed.

\begin{figure}
\includegraphics[width=0.45\textwidth]{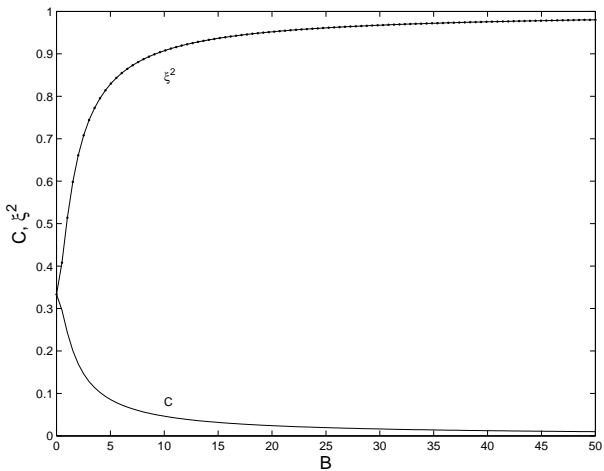}
\caption{The concurrence and spin squeezing parameter versus the
transverse field  for the ground state of TIM ($J=1$).}
\end{figure}

Analytically,
from Eqs.~(\ref{ccc}) and (\ref{xixixi}), we obtain a relation between the concurrence and the squeezing parameter
\begin{equation}
C=\frac{|\xi^2-1|}{2}=\left\{
\begin{array}{ll}
\frac{1-\xi^2}{2} & \text{if} ~~\xi^2\le 1, \\
\frac{\xi^2-1}{2} & \text{if} ~~\xi^2>1.
\end{array}
\right..\label{relation}
\end{equation}
For $\xi^2<=1$, we recover the relation~\cite{WangBarry}.
\begin{equation}
\xi^2=1-2C.
\end{equation}
For $\xi^2>1$, the squeezing parameter and the concurrence satisfy
an interesting relation
\begin{equation}
\xi^2=1+2C.
\end{equation}
Specifically, for the W state, $C=2/3$, and from the relation, we
have $\xi^2=7/3$. Having studied the entanglement and squeezing
properties of the general superposition state of $|111\rangle$ and
W state, we next investigate the ground state of the TIM.

Substituting the expression of $a_1$ and $a_2$ in Eq.~(\ref{gs})
to Eqs.~(\ref{ccc}) and (\ref{xixixi}), we then obtain the
analytical expressions of concurrence and the squeezing parameter
for the ground state. Figure 2 gives the numerical results of the
concurrence and squeezing parameter versus the transverse field
$B$. The spin squeezing parameter monotonously increases with the
increase of the magnetic field, while the concurrence monotonously
decreases. In the limit of $B\rightarrow\infty$, the ground state
will be $|111\rangle$, and there are no spin squeezing ($\xi^2=1$)
and no pairwise entanglement ($C=0$). It is also interesting to
consider another limit of $B\rightarrow 0$, in which the
coefficients $a_1\rightarrow \sqrt{3}/2$ and $a_2\rightarrow
-1/2$. Thus, from Eqs.~(\ref{ccc}) and (\ref{xixixi}), the
concurrence and the spin squeezing parameter approach to 1/3 in
the limit.

We also observe from Fig.~2 that the squeezing parameter $\xi^2\le
1$ for any $B>0$. Thus, from relation (\ref{relation}), we know
that the entanglement and spin squeezing are equivalent for the
ground state. In fact, from Eqs.~(\ref{gs}) and (\ref{xixixi}),
the inequality $\xi^2\le 1$ can be easily proved for the ground
state

{\em Generation of the W state.}  The ground state is a
superposition of W state and state $|111\rangle$, implying that it
is possible to generate W state in this system. We now consider
the Hamiltonian dynamics to generate the W statem.

Let the initial state of the system be
\begin{equation}
|\psi(0)\rangle =|111\rangle.
\end{equation}
Then, the dynamics of the system will happen in the
two-dimensional subspace spanned by $|111\rangle$ and the W state.
The Hamiltonian $H$ in this subspace is expressed as
\begin{align}
H=&\left(
\begin{array}{ll}
-3B&\sqrt{3}J\\
\sqrt{3}J&2J+B
\end{array}
\right)\nonumber\\
=&J-B-(J+2B)\sigma_z+\sqrt{3}J\sigma_x. \label{HHH}
\end{align}
The action of the evolution operator
\begin{equation}
U=\exp(-iHt)
\end{equation}
leads to oscillations between state $|111\rangle$ and the W state.
From Eq.~(\ref{HHH}), the state vector at time $t$, up to a global
phase factor, is obtained as
\begin{align}
|\psi (t)\rangle =&\left[ \cos \omega t+i\cos \theta \sin \omega t\right] |111\rangle, \nonumber\\
&+i\sin \theta \sin \omega t|W\rangle\nonumber\\
\omega=&\sqrt{3J^2+(J+2B)^2},\nonumber\\
\theta =&\arctan[-\sqrt{3}J/(J+2B)],
\end{align}
Specifically, when $J+2B$=0 and $t=\pi/(2\sqrt{3}|J|)$, the state
vector is just the W state. Note that we have removed the previous
assumption $J>0$ and $B>0$ here. In order to generate the W state,
$J$ and $B$ must have opposite signs and satisfy $|J|=2|B|$. We
see that the entangled W state can be exactly generated by
appropriate choosing parameters and initial states. In the scheme
of generating the W state given in Ref.~\cite{Wang01abc}, it is
crucial that there must have one excitation in the initial state.
However, in the present scheme, no excitation is needed for the
initial state.

In conclusion, we have first obtained the exact ground state of
the three-qubit TIM, and then the analytical results of
entanglement and spin squeezing. It has been found that the
pairwise entanglement and spin squeezing are equivalent for the
ground state of the three-qubit TIM. We have proposed a scheme for
generating three-qubit W state via Hamiltonian evolution. There
are many physical systems which can be described by the TIM, and
the scheme could be feasible in experiments.

\acknowledgements X. Wang thanks for the valuable discussions with
Allan I Solomon and HongChen Fu.

\end{document}